\documentclass[aps,twocolumn,10pt,pra,superscriptaddress,showkeys]{revtex4-2}
\usepackage{amsfonts}
\usepackage{amsmath}
\usepackage{amssymb}
\usepackage{mathtools}
\usepackage{algorithm}
\usepackage{algorithmic}
\usepackage{graphicx}
\usepackage{xcolor}
\usepackage{adjustbox}
\usepackage{array}
\usepackage{csquotes}
\usepackage{subcaption}
\usepackage[normalem]{ulem}

\usepackage[colorlinks = true, linkcolor = blue, urlcolor=blue, citecolor =
red, anchorcolor = green,]{hyperref}%
\newtheorem{theorem}{Theorem}
\newtheorem{remark}[theorem]{Remark}
\allowdisplaybreaks

\newcolumntype{P}[1]{>{\centering\arraybackslash}p{#1}}

\newcommand\vqedpsandwich{\adjustbox{valign=m, vspace=0.0pt}{\includegraphics[width=.30\linewidth]{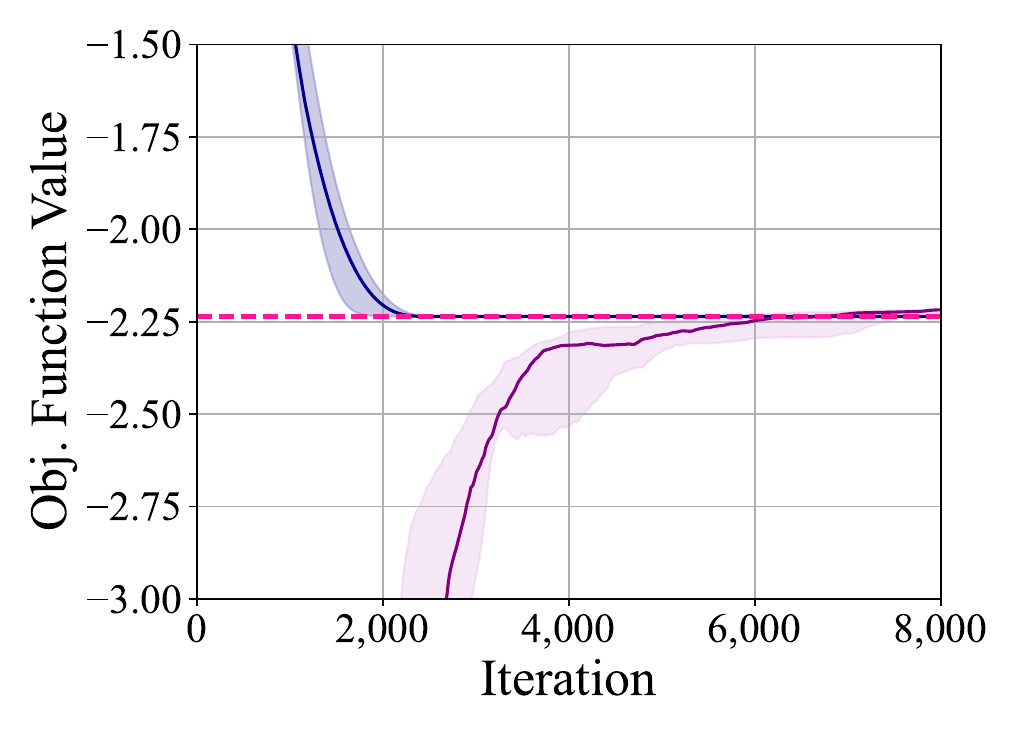}}}
\newcommand\vqedperror{\adjustbox{valign=m, vspace=0.0pt}{\includegraphics[width=.30\linewidth]{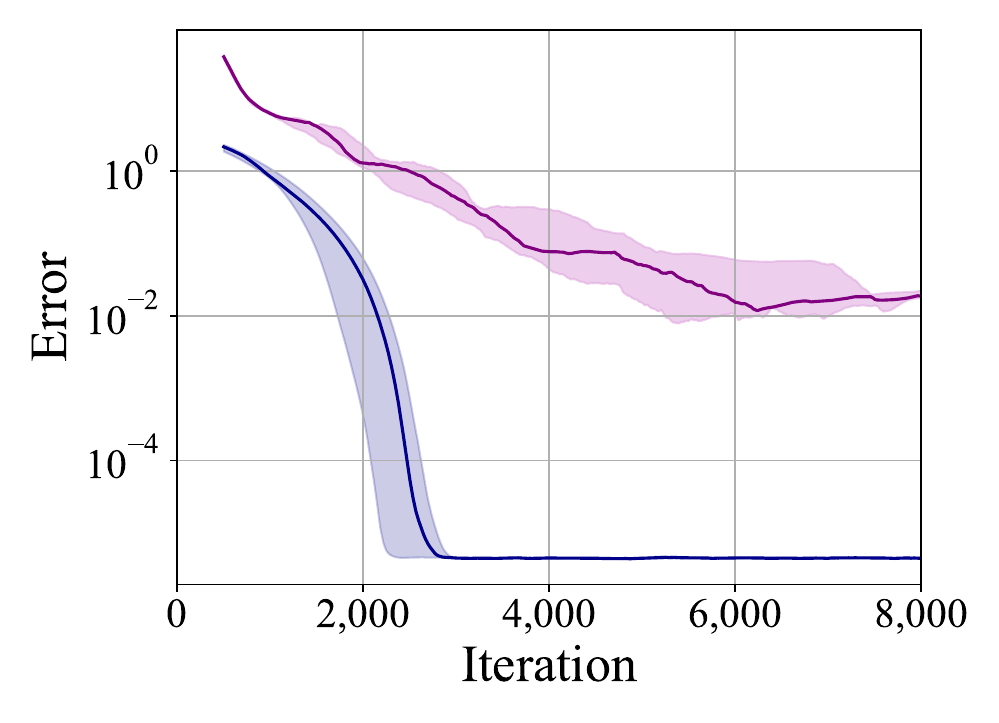}}}
\newcommand\vqedppenalty{\adjustbox{valign=m, vspace=0.0pt}{\includegraphics[width=.30\linewidth]{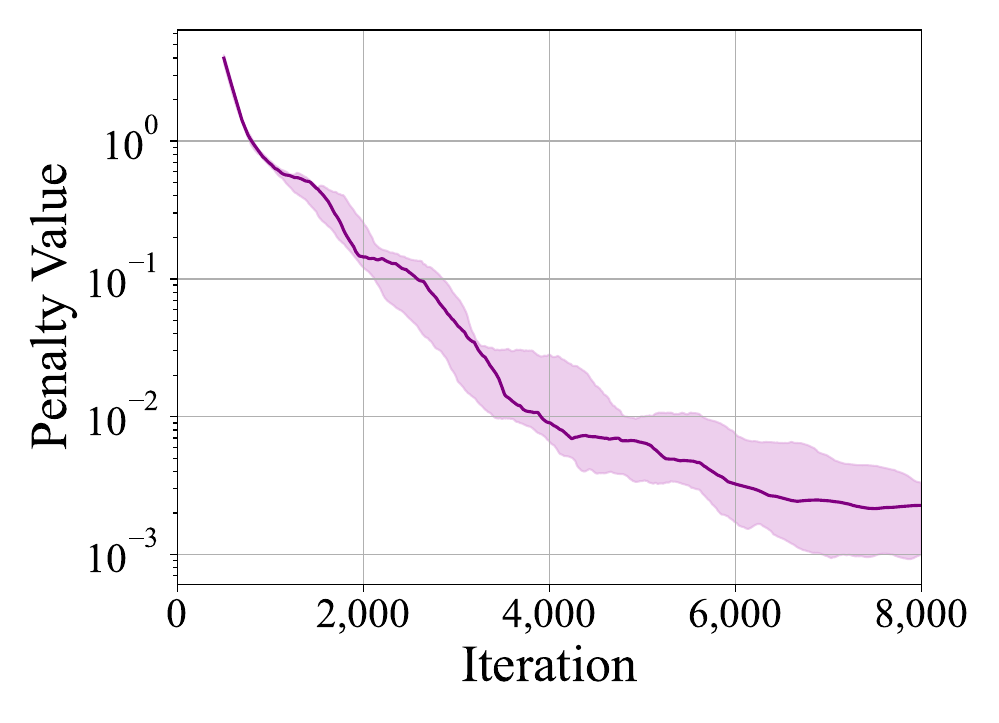}}}

\newcommand\vqedccasandwich{\adjustbox{valign=m, vspace=0.0pt}{\includegraphics[width=.30\linewidth]{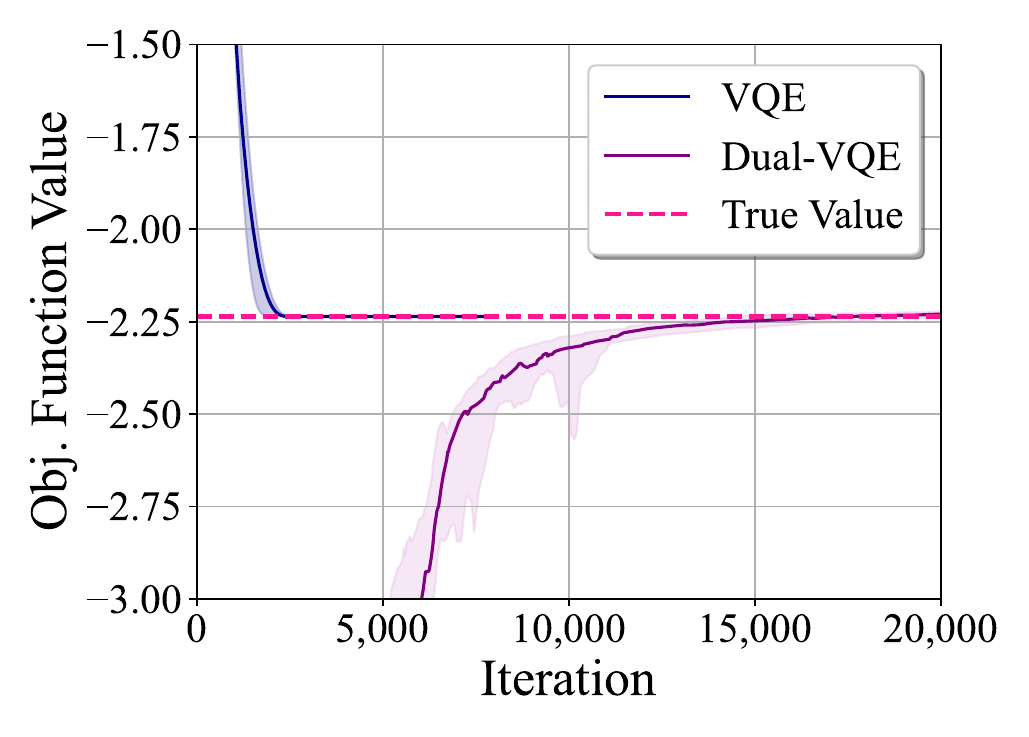}}}
\newcommand\vqedccaerror{\adjustbox{valign=m, vspace=0.0pt}{\includegraphics[width=.30\linewidth]{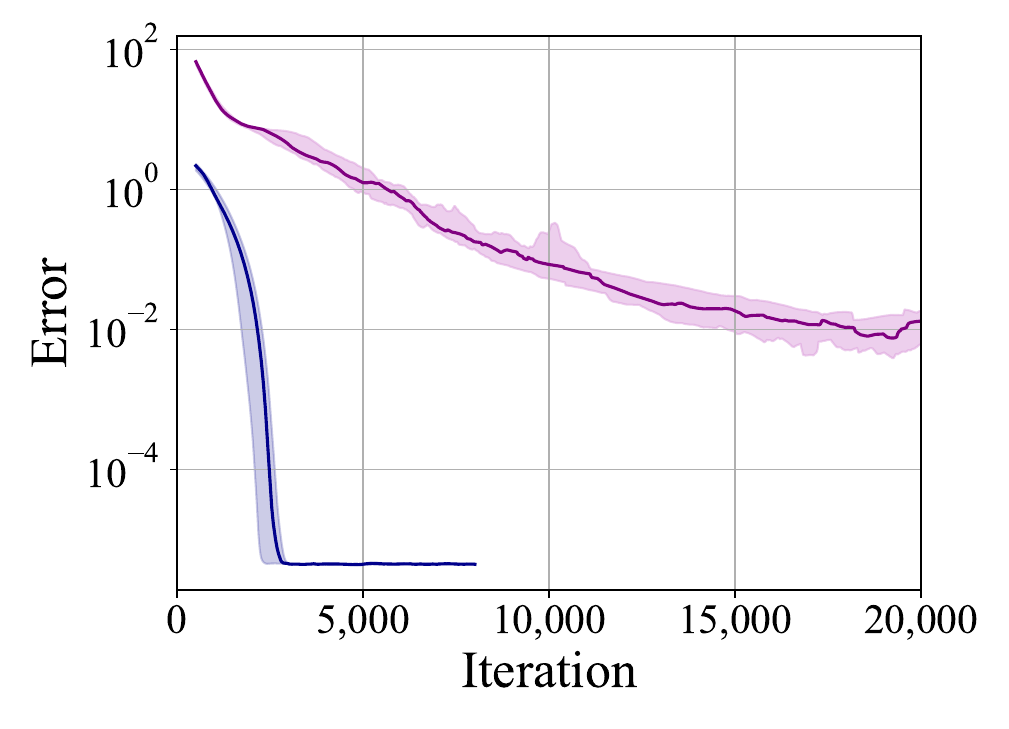}}}
\newcommand\vqedccapenalty{\adjustbox{valign=m, vspace=0.0pt}{\includegraphics[width=.30\linewidth]{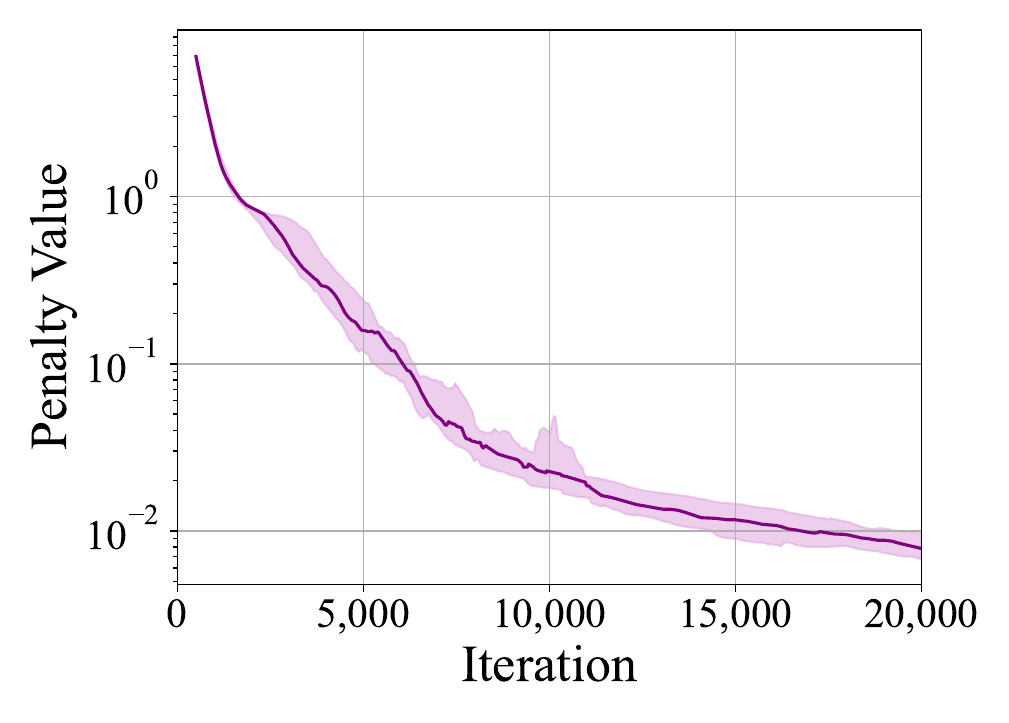}}}

\begin{document}
\preprint{ }
\title[QSlack]{Dual variational quantum eigensolver: A quantum algorithm to lower bound the ground-state energy }
\author{Hanna Westerheim}
\affiliation{School of Applied and Engineering Physics, Cornell University, Ithaca, New
York 14850, USA}
\affiliation{School of Electrical and Computer Engineering, Cornell University, Ithaca, New
York 14850, USA}
\author{Jingxuan Chen}
\affiliation{Department of Computer Science, Cornell University, Ithaca, New York 14850, USA}
\affiliation{School of Electrical and Computer Engineering, Cornell University, Ithaca, New
York 14850, USA}
\author{Zo\"e Holmes}
\affiliation{Institute of Physics, Ecole Polytechnique F\'ed\'erale de Lausanne (EPFL),
CH-1015 Lausanne, Switzerland}
\author{Ivy Luo}
\affiliation{School of Electrical and Computer Engineering, Cornell University, Ithaca, New
York 14850, USA}
\author{Theshani Nuradha}
\affiliation{School of Electrical and Computer Engineering, Cornell University, Ithaca, New
York 14850, USA}
\author{Dhrumil Patel}
\affiliation{Department of Computer Science, Cornell University, Ithaca, New York 14850, USA}
\author{Soorya Rethinasamy}
\affiliation{School of Applied and Engineering Physics, Cornell University, Ithaca, New
York 14850, USA}
\author{Kathie Wang}
\affiliation{School of Electrical and Computer Engineering, Cornell University, Ithaca, New
York 14850, USA}
\author{Mark M.~Wilde}
\affiliation{School of Electrical and Computer Engineering, Cornell University, Ithaca, New
York 14850, USA}
\begin{abstract}
The variational quantum eigensolver (VQE) is a hybrid quantum--classical variational algorithm that produces an upper-bound estimate of the ground-state energy of a Hamiltonian. As quantum computers become more powerful and go beyond the reach of classical brute-force simulation, it is important to assess the quality of solutions produced by them. Here we propose a dual variational quantum eigensolver (dual-VQE) that produces a lower-bound estimate of the ground-state energy. As such, VQE and dual-VQE can serve as quality checks on their solutions; in the ideal case, the VQE upper bound and the dual-VQE lower bound form an interval containing the true optimal value of the ground-state energy. The idea behind dual-VQE is to employ semidefinite programming duality to rewrite the ground-state optimization problem as a constrained maximization problem, which itself can be bounded from below by an unconstrained optimization problem to be solved by a variational quantum algorithm. When using a convex combination ansatz in conjunction with a classical generative model, the quantum computational resources needed to evaluate the objective function of dual-VQE are no greater than those needed for that of VQE. We also show that the problem is well suited for classical pretraining using matrix product states and these methods help warm-start the optimization. We simulated the performance of dual-VQE on the transverse-field Ising model with and without pretraining and found that, for the example considered, while dual-VQE training is slower and noisier than VQE, it approaches the true value with an error of order~$10^{-2}$.
\end{abstract}
\date{\today}
\maketitle

\tableofcontents

\section{Introduction}

\subsection{Background}

A key approach hoped to achieve quantum advantage on near-term quantum computers involves the use of variational quantum algorithms~\cite{bharti2021noisy,Cerezo2021}.
The variational quantum eigensolver (VQE) was the first such
algorithm proposed~\cite{Peruzzo2014}: it estimates an upper bound on the
ground-state energy of a Hamiltonian. The basic idea behind it is simple,
based on the well known variational principle from physics~\cite{Gerjuoy1983}.
For a Hamiltonian $H$, its ground-state energy is equal to its minimum
eigenvalue $\lambda_{\min}(H)$, which is equal to the following minimization
problem:
\begin{equation}
\lambda_{\min}(H)   =\inf_{\rho\in\mathcal{D}}\operatorname{Tr}
[H\rho]\label{eq:SDP-VQE}
\end{equation}
where $\mathcal{D}$ is the set of density matrices (positive semidefinite
matrices with trace equal to one). Since the objective function
$\operatorname{Tr}[H\rho]$ is linear in $\rho$ and the set of density matrices
is convex, the optimal value of~\eqref{eq:SDP-VQE} is obtained on the boundary of the set $\mathcal{D}$ (i.e., the set of pure states). It then follows that
\begin{equation}
    \lambda_{\min}(H) = \inf_{\vert \psi \rangle \in \mathcal{P}} \langle\psi |H| \psi \rangle,
    \label{eq:min-H-pure}
\end{equation}
where $\mathcal{P}$ is the set of state vectors.
In a variational approach, the set of pure states is approximated by a parametrized set $\{\vert \psi(\theta) \rangle\}_{\theta \in \Theta}$ of states, where $\Theta$ is a parameter set and $\theta$ is a parameter vector. The equality in~\eqref{eq:min-H-pure} is preserved if the set of parametrized pure states contains an optimal pure state. In reality, guaranteeing this condition is not possible, leading to the following upper bound on the ground-state energy:
\begin{equation}
\lambda_{\min}(H)\leq\inf_{\theta\in\Theta}\langle\psi(\theta)|H|\psi
(\theta)\rangle. \label{eq:var-principle}%
\end{equation}

An example of wide interest is when $n\in\mathbb{N}$ and $H$ is a $2^{n}%
\times2^{n}$ Hermitian matrix, describing the local interactions that take
place between $n$ spin-1/2 particles. In this case, $H=\sum_{j=1}^{L}H_{j}$,
where each $H_{j}$ is a Hermitian matrix of constant size that acts
nontrivially on a constant number of particles. Observing that the
optimization in~\eqref{eq:SDP-VQE} is a semidefinite program (SDP)
\cite{Skrzypczyk2023}, one could attempt to solve it by means of well known
approaches used to solve SDPs; however, the computational complexity of such
approaches is exponential in~$n$ and thus infeasible. Standard classical
approaches for solving~\eqref{eq:SDP-VQE} in the general case suffer from the
same problem, or they avoid it by not allowing for the optimization to include
highly entangled states (see~\cite{Echenique2007} for a review of some of these methods).

VQE\ instead parametrizes the set of state vectors by a parametrized quantum
circuit $U(\theta)$ acting on an easily preparable initial state $|\psi
_{0}\rangle$, leading to $|\psi(\theta)\rangle\coloneqq U(\theta)|\psi
_{0}\rangle$, and measures the expectation of each observable $H_{j}$ with
respect to $|\psi(\theta)\rangle$ in order to obtain an estimate
$\widehat{E_{j}}$ of $\langle\psi(\theta)|H_{j}|\psi(\theta)\rangle$. Then the
sum $\sum_{j}\widehat{E_{j}}$ is an estimate of the desired expectation
$\langle\psi(\theta)|H|\psi(\theta)\rangle$.
The estimates of $\langle\psi
(\theta)|H|\psi(\theta)\rangle$ and its gradient can be fed to a classical
optimization procedure, and this process iterated until convergence or a
maximum number of iterations is reached.
As a consequence of
\eqref{eq:var-principle}, this procedure in the ideal case produces an
estimate of an upper bound on the ground-state energy $\lambda_{\min}(H)$.

\subsection{Main contribution: Dual-VQE method}

In the short term, while classical computers are still competitive with
small-scale quantum computers, it is possible to test how well VQE\ is
performing by comparing its value with an extremely precise estimate of the
true known value. As mentioned above, an SDP\ can output such a precise
estimate for sufficiently small system sizes. However, this approach is not
viable in the long term, which leads to the key question that our paper
addresses: 
\begin{displayquote}
\textit{How can we assess the quality of VQE's upper-bound
estimate of $\lambda_{\min}(H)$ without access to a reliable classical
algorithm for estimating $\lambda_{\min}(H)$?}
\end{displayquote}

We address this question by developing a variational quantum algorithm (called
dual-VQE) to estimate a lower bound on the ground-state energy, which can be
compared to the upper-bound estimate obtained from VQE. Since the two
estimates ideally sandwich the true optimal value, they serve as quality
checks on each other. The key ideas of our approach are to 1)\ use
SDP\ duality theory to reformulate $\lambda_{\min}(H)$ as a maximization
problem and 2) further reformulate this maximization problem in such a way
that it can be estimated by means of a variational quantum algorithm, which
finally leads to a lower-bound estimate of $\lambda_{\min}(H)$. We call our
method dual-VQE because it makes use of the dual characterization of
$\lambda_{\min}(H)$ to arrive at this lower-bound estimate. After executing
VQE\ and dual-VQE, one can then compare the upper-bound and lower-bound
estimates to understand how well a quantum computer can approximate the true
optimal value $\lambda_{\min}(H)$, without the aid of precise classical
methods.

The dual-VQE\ method involves 1)\ converting inequality
constraints to equality constraints by means of positive semidefinite slack
variables, 2)\ converting a constrained optimization into an unconstrained one
by means of a penalty term in the objective function, 3)\ replacing positive
semidefinite slack variables with scaled parametrized quantum states that
are efficiently preparable on quantum computers, and 4)\ estimating various
terms in the objective function by means of the destructive swap test
\cite{GC13} and sampling estimates of expectations of observables. Our
companion paper~\cite{Chen2023qslack} explores this general approach in a much broader
context beyond the ground-state energy problem.

Although there have been various classical algorithms proposed for obtaining
lower bounds on the ground-state energy
\cite{Mazziotti2004,Mazziotti2004a,Mazziotti2006,Barthel2012,haim2020variationalcorrelations}%
, they all ultimately suffer from the exponential increase of the space on
which $H$ acts, as the number of particles increases. Our approach is
fundamentally different, involving a reformulation of the original
optimization problem in terms of its SDP dual, and then bounding that quantity
from below using a variational quantum algorithm. Our approach is also complementary to~\cite[Section~VI-A]{BHVK22}, which does not take a variational approach, nor does it provably obtain a lower bound on the ground-state energy.

\subsection{Pretraining with matrix product states}

A key issue that plagues variational quantum algorithms is the presence of barren plateaus in the training landscape~\cite{McClean_2018, fontana2023adjoint,ragone2023unified}, meaning that the gradient (or equivalently cost differences~\cite{arrasmith2022equivalence}) vanishes exponentially fast in the number of qubits. At their core, barren plateaus arise from a \textit{curse of dimensionality}~\cite{cerezo2023does} stemming from the exponentially scaling size of the Hilbert space. On the other hand, standard techniques for provably avoiding barren plateaus do so by encoding the relevant dynamics in a smaller subspace making it effectively possible to classically simulate the parametrized quantum circuit~\cite{cerezo2023does}. 

An important question for dual-VQE, shared also with other variational quantum algorithms, is thus whether we can identify quantum models that both avoid barren plateaus and cannot be classically simulated.
A promising avenue to do so is to note that theoretical analysis of barren plateaus and the simulability of variational quantum algorithms is confined to average case analyses. It is possible that there might exist clever initialization strategies to explore small subsections of the landscape that exhibit substantial gradients, and are also hard to simulate classically.

Motivated as such, in this work we develop a matrix product state (MPS) pretrainer that uses tensor networks to initialize the ansatz parameters in a more efficient way~\cite{HPM+19, RMM+23}. 
We find that this MPS pretrainer allows us to start the quantum optimization from a favorable initial point, compared to random initializations, as evidenced by the convergence of the cost function value closer to the true value (in this context, see Figure~\ref{fig:pretraining-plots}).

We find that, without pretraining, the dual-VQE approach on a two-qubit example takes more iterations to converge as compared to the standard VQE optimization; however, it approaches the true value with error on the order of $10^{-2}$. We also note that in this example, the objective function values across training are noisier for dual-VQE when compared to VQE. We expect this behavior due to the increase in number of terms and parameters when estimating the ground-state energy with dual-VQE. Details of numerical simulations are given below, and training plots are shown in Figure~\ref{fig:vqed-plots}. 

With the introduction of pretraining, we tackle a larger problem and find that, as expected, pretraining with a larger bond dimension allows for more efficient optimization. Concretely, with a maximum bond dimension of eight, we find the ground state of a three-qubit model with a relative error of $0.5\%$.

\subsection{Paper organization}

The rest of our paper proceeds by detailing the dual-VQE method outlined above. In Section~\ref{sec:theory}, we first describe the theoretical derivation of the key result. We then follow this up in Section~\ref{sec:alg-cons} with more intricate details of the simulations, including the penalty value, which plays a key role in the theoretical derivation, the ansatz choices, and the pretraining method that we employ. Finally, in Section~\ref{sec:Results}, we report the results of simulations of the transverse-field Ising model, a key example of physical interest. Appendix~\ref{app:matrix-product-states} provides background on matrix product states, and Appendix~\ref{app:alg_AD_OD} provides further details of the pretraining method that we used. 

\section{Theoretical derivation}
\label{sec:theory}

Let us first apply basic SDP duality theory to derive a well known result, that the following strong-duality
equality holds for the optimization problem in~\eqref{eq:SDP-VQE}:
\begin{equation}
\sup_{\eta\in\mathbb{R}}\left\{  \eta:\eta I\leq H\right\}  =\inf_{\rho
\in\mathcal{D}}\operatorname{Tr}[H\rho], \label{eq:SDP-dual-lam-min}%
\end{equation}
where $\sup_{\eta\in\mathbb{R}}\left\{  \eta:\eta I\leq H\right\}  $ is the
SDP dual of~\eqref{eq:SDP-VQE}. 
To see this, recall that the
matrix inequality $A\geq B$ holds for Hermitian matrices $A$ and $B$ if $A-B$
is a positive semidefinite matrix. Then consider that
\begin{align}
\inf_{\rho\in\mathcal{D}}\operatorname{Tr}[H\rho]  &  =\inf_{\rho\geq
0}\left\{  \operatorname{Tr}[H\rho]:\operatorname{Tr}[\rho]=1\right\} \\
&  =\inf_{\rho\geq0}\sup_{\eta\in\mathbb{R}}\left\{  \operatorname{Tr}%
[H\rho]+\eta\left(  1-\operatorname{Tr}[\rho]\right)  \right\} \\
&  =\inf_{\rho\geq0}\sup_{\eta\in\mathbb{R}}\left\{  \eta+\operatorname{Tr}%
[\left(  H-\eta I\right)  \rho]\right\} \\
&  \geq\sup_{\eta\in\mathbb{R}}\inf_{\rho\geq0}\left\{  \eta+\operatorname{Tr}%
[\left(  H-\eta I\right)  \rho]\right\} \\
&  =\sup_{\eta\in\mathbb{R}}\left\{  \eta:\eta I\leq H\right\}  .
\end{align}
The second equality follows by introducing the Lagrange multiplier $\eta$ and
because the constraint $\operatorname{Tr}[\rho]=1$ does not hold if and only
if $\sup_{\eta\in\mathbb{R}}\left\{  \eta\left(  1-\operatorname{Tr}%
[\rho]\right)  \right\}  =+\infty$. The third equality holds by simple
algebra. The sole inequality follows from the standard max-min inequality. The
final equality follows by thinking of $\rho$ as a matrix Lagrange multiplier
and because the constraint $\eta I\leq H$ does not hold if and only if
$\inf_{\rho\geq0}\left\{  \operatorname{Tr}[\left(  H-\eta I\right)
\rho]\right\}  =-\infty$. Thus, we have established the weak-duality inequality \begin{equation}
    \sup_{\eta\in\mathbb{R}}\left\{  \eta:\eta I\leq H\right\}  \leq \inf_{\rho
\in\mathcal{D}}\operatorname{Tr}[H\rho].
\end{equation} 
Note that  equality holds in
\eqref{eq:SDP-dual-lam-min} due to Slater's theorem~\cite{BV04}: we can pick
$\rho=I/d$, where $d$ is the dimension of the underlying Hilbert space, and
$\eta<\lambda_{\min}(H)$ as strictly feasible choices in both the dual and
primal, thus satisfying the conditions of Slater's theorem. 

Now that we have recalled this standard result, let us focus on showing how to
estimate the left-hand side of~\eqref{eq:SDP-dual-lam-min} on a quantum computer.
Recalling that the matrix inequality $\eta I\leq H$ is equivalent to the
existence of a positive semidefinite matrix $W$ such that $H-\eta I=W$, we
conclude that%
\begin{equation}
\sup_{\eta\in\mathbb{R}}\left\{  \eta:\eta I\leq H\right\}  =\sup_{\eta
\in\mathbb{R},W\geq0}\left\{  \eta:H-\eta I=W\right\}  .
\end{equation}
The matrix $W$ is known in the theory of optimization~\cite{BV04} as a slack
variable because it eliminates the slack in the inequality constraint $\eta
I\leq H$ by reducing it to an equality constraint.

Our next observation, related to observations used in~\cite{patel2021variational}, is
that every positive semidefinite matrix $W\neq0$ can be written as a scaled
density matrix, i.e., $W=\nu\omega$, where $\nu=\operatorname{Tr}[W]$ and
$\omega=W/\nu\in\mathcal{D}$, implying that
\begin{equation}
\sup_{\substack{\eta\in\mathbb{R},\\W\geq0}}\left\{  \eta:H-\eta I=W\right\}
=\sup_{\substack{\eta\in\mathbb{R},\nu\geq0,\\\omega\in\mathcal{D}}}\left\{
\eta:H-\eta I=\nu\omega\right\}  .
\end{equation}

Next, we adopt a standard approach of introducing a penalty term in the
objective function to transform the constrained optimization to an
unconstrained one:%
\begin{multline}
\sup_{\substack{\eta\in\mathbb{R},\nu\geq0,\\\omega\in\mathcal{D}}}\left\{
\eta:H-\eta I=\nu\omega\right\} \\
= \lim_{c\to \infty} \sup_{\substack{\eta\in\mathbb{R},\nu\geq
0,\\\omega\in\mathcal{D}}}\left\{  \eta-c\left\Vert H-\eta I-\nu
\omega\right\Vert _{2}^{2}\right\}  , \label{eq:penalty-method-dual-VQE}%
\end{multline}
where $c >0 $ is a penalty parameter and $\left\Vert Z\right\Vert _{2}%
\coloneqq\sqrt{\operatorname{Tr}[Z^{\dag}Z]}$ is the Hilbert--Schmidt norm of
a matrix $Z$. Note that the equality in~\eqref{eq:penalty-method-dual-VQE}  holds whenever the penalty term is faithful
(specifically, we invoke~\cite[Proposition~5.2.1]{Bertsekas2016} to justify
this conclusion, setting $\lambda_{k}=0$ therein for all
$k$). By faithfulness, we mean that the
penalty term is equal to zero if and only if $H-\eta I=\nu\omega$. This is
true for the Hilbert--Schmidt norm, and this particular norm has the
convenient advantage that it can be estimated by quantum sampling procedures
that we recall below.

The final step is to modify the optimization in~\eqref{eq:penalty-method-dual-VQE} to be over a
parametrized set $\left\{  \omega(\theta)\right\}  _{\theta\in\Theta}$ of
density matrices instead of being over all density matrices, which implies
that%
\begin{multline}
\sup_{\substack{\eta\in\mathbb{R},\nu\geq0,\\\theta\in\Theta}}\left\{
\eta-c\left\Vert H-\eta I-\nu\omega(\theta)\right\Vert _{2}^{2}\right\}
\label{eq:parameter-set-lower-bnd}\\
\leq\sup_{\substack{\eta\in\mathbb{R},\nu\geq0,\\\omega\in\mathcal{D}%
}}\left\{  \eta-c\left\Vert H-\eta I-\nu\omega\right\Vert _{2}^{2}\right\}  .
\end{multline}

Piecing together all these steps in~\eqref{eq:SDP-dual-lam-min}--\eqref{eq:parameter-set-lower-bnd},  expanding the Hilbert--Schmidt norm, and defining
\begin{equation}
f^{\ast}(c)\coloneqq\sup_{\eta\in\mathbb{R},\nu\geq0,\theta\in\Theta}%
f(\eta,\nu,\theta,c),
\label{eq:f-c-optimized}
\end{equation}
where $\Theta$ is a parameter set,
\begin{multline}
f(\eta,\nu,\theta,c)\coloneqq\eta-c\big(\operatorname{Tr}[H^{2}]+\eta^{2}%
2^{n}+\nu^{2}\operatorname{Tr}[\omega(\theta)^{2}]\\
-2\eta\operatorname{Tr}[H]-2\nu\operatorname{Tr}[H\omega(\theta)]+2\eta
\nu\big)\, ,
\label{eq:dual-vqe-objective}
\end{multline}
and $c$ is a penalty parameter,
we obtain the main theoretical result supporting the dual-VQE approach:

\begin{theorem}[Dual-VQE lower bound]
\label{thm:dVQE-lower-bound}
The function $f^{\ast}(c)$ in~\eqref{eq:f-c-optimized} provides a lower bound on the ground-state energy of the Hamiltonian~$H$ in the limit $c\to \infty$, i.e.,
\begin{equation}
\lim_{c\to\infty} f^{\ast}(c)\leq\lambda_{\min}(H) .
\label{eq:key-result-dual-vqe}%
\end{equation}
\end{theorem}

In the rest of the work, we use the term ``unpenalized objective function'' to refer to $\eta$, ``penalty'' to refer to the sum of all terms in~\eqref{eq:dual-vqe-objective} that are multiplied by the penalty parameter $c$, i.e., 
\begin{multline}
    \operatorname{Tr}[H^{2}]+\eta^{2}%
    2^{n}+\nu^{2}\operatorname{Tr}[\omega(\theta)^{2}]\\
    -2\eta\operatorname{Tr}[H]-2\nu\operatorname{Tr}[H\omega(\theta)]+2\eta
    \nu\, ,
\end{multline}
and ``objective function'' to refer to $f$ in \eqref{eq:dual-vqe-objective} (i.e., the unpenalized objective function minus $c$ times the penalty).

\section{Algorithm construction}

\label{sec:alg-cons}

The basic idea behind the dual-VQE\ algorithm is to  follow the standard penalty method approximately~\cite[Section~5.2]{Bertsekas2016}:\ define a sequence $\left(  f^{\ast}(c_{k}) \right)_{k \in \mathbb{N}}$ of optimization problems where $\left(c_{k}\right) _{k\in\mathbb{N}}$ is a monotone increasing sequence of penalty parameters. We then solve the optimization for each $k$ and use the solution as the initial guess for the next iteration. Under the assumption that each optimization $f^{\ast}(c_{k})$ can be solved, the method is guaranteed to converge to the correct solution. In our simulations, we use a variety of update methods to change the value of $c$ as the optimization progresses. In some cases, we pick $c$ to be a high value and keep it constant throughout the optimization. In other cases, we begin with a smaller value, and ramp it up as the iteration count grows.

Inspecting~\eqref{eq:dual-vqe-objective}, only two terms $\operatorname{Tr}[\omega(\theta)^{2}]$ and $\operatorname{Tr}[H\omega(\theta)]$
depend on the optimization variable $\omega(\theta)$, while constant terms like
$\operatorname{Tr}[H^{2}]$ and $\operatorname{Tr}[H]$ (having no dependence on
$\eta$, $\nu$, or $\omega(\theta)$) can be calculated offline.



A key case of interest for physics and chemistry is when~$H$ is a linear
combination of Pauli strings, so that%
\begin{equation}
H=\sum_{\overrightarrow{x}}\alpha_{\overrightarrow{x}}\sigma_{\overrightarrow
{x}},
\end{equation}
where $\overrightarrow{x}\in\left\{  0,1,2,3\right\}  ^{n}$, $\alpha
_{\overrightarrow{x}}\in\mathbb{R}$, $\sigma_{\overrightarrow{x}}\equiv
\sigma_{x_{1}}\otimes\cdots\otimes\sigma_{x_{n}}$ is a Pauli string, and
$\sigma_{0}\equiv I$, $\sigma_{1}\equiv\sigma_{X}$, $\sigma_{2}\equiv
\sigma_{Y}$, and $\sigma_{3}\equiv\sigma_{Z}$ are the standard Pauli matrices.
For this case, it follows from substitution and the orthogonality relation
$\operatorname{Tr}[\sigma_{\overrightarrow{x}}\sigma_{\overrightarrow{y}%
}]=2^{n}\delta_{\overrightarrow{x},\overrightarrow{y}}$ that%
\begin{align}
\operatorname{Tr}[H^{2}]  &  =2^{n}\left\Vert \overrightarrow{\alpha
}\right\Vert _{2}^{2},\\
\operatorname{Tr}[H]  &  =2^{n}\alpha_{\overrightarrow{0}},\\
\operatorname{Tr}[H\omega(\theta)]  &  =\sum_{\overrightarrow{x}}\alpha
_{\overrightarrow{x}}\operatorname{Tr}[\sigma_{\overrightarrow{x}}\omega(\theta)].
\end{align}
As is common in  Hamiltonian models relevant for physics and chemistry, the number of
nonzero coefficients in the tuple $\left(  \alpha_{\overrightarrow{x}
}\right)  _{\overrightarrow{x}}$ is polynomial in$~n$, so that the objective
function in~\eqref{eq:dual-vqe-objective} can be efficiently estimated. In this work, we focus on the transverse-field Ising model given in~\eqref{eq:ising_model}, which falls under this category. 

In dual-VQE, we estimate the
expectation $\operatorname{Tr}[H\omega(\theta)]$ by sampling from a quantum
computer, and we use other approaches, either quantum or classical, for
estimating the purity $P \equiv \operatorname{Tr}\!\left[\left(  \omega(\theta)\right)  ^{2}
\right]$. If employing gradient descent for solving each optimization, one should estimate the gradient as well, which we discuss further below. The main additional overhead compared to VQE\ is the need to employ an ansatz for generating a mixed state $\omega(\theta)$ and to estimate the purity~$P$, given that VQE\ already estimates a term like $\operatorname{Tr}[H\omega(\theta)]$. However, as we discuss below, one approach reduces the estimation of the purity~$P$ to a classical sampling task, thus removing the need for a quantum computer for this part.

\subsection{Parametrization}

There are at least two general ways of parametrizing the set of mixed states,
called the purification ansatz~\cite{CSZW20,patel2021variational,Ezzell2023} and the
convex combination ansatz
\cite{verdon2019quantum,Liu2021,Ezzell2023,sbahi2022provably} (see also~\cite{Chen2023qslack}). Focusing on the first
one to start, it is based on the purification principle
\cite{bures1969extension,Uhlmann1976,Uhlmann1986}, in which every mixed state
is realized as the marginal of a pure state on a larger system. For the
purification ansatz, we simply start with a parametrized family $\left\{
|\psi(\theta)\rangle_{RS}\right\}  _{\theta\in\Theta}$ of pure states of a
reference system~$R$ and the system $S$ of interest, and then we obtain a
parametrized family $\left\{  \omega(\theta)\right\}  _{\theta\in\Theta}$ of
mixed states via the partial trace $\omega(\theta)=\operatorname{Tr}_{R}
[|\psi(\theta)\rangle\!\langle\psi(\theta)|_{RS}]$. We can estimate the purity
$\operatorname{Tr}[\left(  \omega(\theta)\right)  ^{2}]$ by means of the
destructive swap test~\cite{GC13} (see, e.g.,~\cite[Section~2.2]
{bandyopadhyay2023efficient} for a precise statement of the algorithm). We can
estimate the gradient of the objective function in
\eqref{eq:key-result-dual-vqe} by means of the parameter-shift rule
\cite{Li2017,Mitarai2018,Schuld2019}, simply because
\begin{equation}
\operatorname{Tr}[H\omega(\theta)]=\langle\psi(\theta)|_{RS}\left(
I_{R}\otimes H_{S}\right)  |\psi(\theta)\rangle_{RS},
\end{equation}
and one can also use a slightly modified version of the parameter-shift rule along with the destructive swap
test to estimate the gradient of the purity term $\operatorname{Tr}
[\omega(\theta)^{2}]$ because
\begin{equation}
\operatorname{Tr}[\omega(\theta)^{2}]=\langle\phi(\theta)|\left(
I_{R_{1}R_{2}}\otimes F_{S_{1}S_{2}}\right)  |\phi(\theta)\rangle,
\label{eq:purity-to-swap}
\end{equation}
where
\begin{equation}
|\phi(\theta)\rangle_{R_{1}R_{2}S_{1}S_{2}}\equiv|\psi(\theta)\rangle
_{R_{1}S_{1}}\otimes|\psi(\theta)\rangle_{R_{2}S_{2}},
\end{equation}
$F$ is the unitary and Hermitian swap operator, and we have suppressed some system labels in
\eqref{eq:purity-to-swap} for brevity (see also~\cite[Eq.~(C2)]{Ezzell2023} in
this context).

The convex combination ansatz is based on the fact that every mixed state can
be written as a convex combination of pure orthogonal states. Following this
idea, we obtain a parametrized family $\left\{  \omega(\varphi,\gamma
)\right\}  _{\varphi,\gamma\in\Theta}$ by separately parametrizing the
eigenvalues and eigenvectors as follows:
\begin{equation}
\omega(\varphi,\gamma)=\sum_{x}p_{\varphi}(x)U(\gamma)|x\rangle\!\langle
x|U(\gamma)^{\dag},
\end{equation}
where $\left(  p_{\varphi}(x)\right)  _{x}$ is a parametrized probability
distribution, $U(\gamma)$ is a parametrized unitary, and $\left\{
|x\rangle\right\}  _{x}$ is the standard computational basis. The parametrized
probability distribution $p_{\varphi}$ can be realized by a
neural-network-based generative model~\cite{Ackley1985,bengio2013estimating,mohamed2020monte} or a quantum
circuit Born machine~\cite{Benedetti2019}, while the parametrized unitary
$U(\gamma)$ can be realized by a parametrized circuit.

Interestingly, although it is necessary to use a quantum computer to estimate
the purity $\operatorname{Tr}[\left(  \omega(\theta)\right)  ^{2}]$ for the
purification ansatz, it is not necessary to do so when using the convex
combination ansatz in conjunction with a neural-network-based generative
model:\ it suffices to employ a classical sampling approach, thus further
reducing the quantum resource requirements for dual-VQE (see also
\cite[Eqs.~(20)--(23)]{Ezzell2023} in this context). This is because
$\operatorname{Tr}[\left(  \omega(\varphi,\gamma)\right)  ^{2}]=\sum
_{x}p_{\varphi}^{2}(x)$, so that one can repeatedly sample from $p_{\varphi}$
and perform a collision test to estimate $\sum_{x}p_{\varphi}^{2}(x)$. Thus,
for the convex combination ansatz used in the aforementioned way, one only
needs a quantum computer to estimate the expectation
\begin{equation}
\operatorname{Tr}[H\omega(\varphi,\gamma)]=\sum_{x}p_{\varphi}(x)\langle
x|U(\gamma)^{\dag}HU(\gamma)|x\rangle,
\end{equation}
which can be done by repeating the following procedure and calculating the
sample mean: take a sample$~x$ from $p_{\varphi}(x)$, prepare the state
$U(\gamma)|x\rangle$, measure this state according to $H$, and record the
outcome. To estimate the gradient of the purity $\sum_{x}p_{\varphi}^{2}(x)$
and the gradient of the expectation $\operatorname{Tr}[H\omega(\varphi
,\gamma)]$ with respect to $\varphi$, perhaps the simplest approach is to use
the simultaneous perturbation stochastic approximation~\cite{Spall1992},
because the distribution $p_{\varphi}(x)$ in this case is not differentiable, and as such, standard methods like backpropagation are not readily applicable (see
\cite{bengio2013estimating} for further discussions and other approaches).

\subsection{Pretraining}

Pretraining broadly refers to a set of computational techniques that are relatively inexpensive and kick start (or warm start) an optimization process, as opposed to a random start. In several domains, pretraining has been shown to be advantageous in reducing the training time and accelerating convergence~\cite{CKA21, MHZ24}. 

In this section, we describe the details of the particular pretraining process used in this work. The overall process consists of two phases: pretraining, followed by a translation into a quantum circuit. The pretraining is conducted using matrix product states (MPS)~\cite{CPD+21}, and the final translation consists of a combination of two techniques, analytic decomposition (AD) and optimizing decomposition (OD), as proposed in~\cite{Ran2020a, Rudolph_2024}. Important preliminaries on tensor networks and matrix product states can be found in Appendix~\ref{app:matrix-product-states}, and key ideas of the pretraining process, including an in-depth review of AD and OD, can be found in Appendix~\ref{app:alg_AD_OD}.

Before delving into the specifics of the algorithm, we aim to establish the rationale behind pretraining using MPS. Our motivation for this approach stems from Refs.~\cite{rudolph2023classical, RMM+23}, which explored solving the Variational Quantum Eigensolver (VQE) problem by pretraining it with MPS. The study demonstrated that parametrized quantum circuits for VQE exhibited significant performance improvements when used in conjunction with a pretrained initialization, in contrast to a random initialization. When applied to the dual-VQE problem, we find that pretraining provides a significant kick start, with the advantage increasing with maximum bond dimension of the MPS. 

The overall process consists of three phases: pretraining, translation, and quantum training. Since it is more complex to map an MPS to a convex combination ansatz state, we used the purification ansatz in our pretraining procedure. We now delve into the three stages in more detail.

\subsubsection{Pretraining using matrix product states}

The dual-VQE problem (without the final parametrization of the state $\omega$) is given by
\begin{equation}
    \label{eq:dvqe-mps-costfn}
\sup_{\substack{\eta\in\mathbb{R},\nu\geq0,\\ \omega \in \mathcal{D}}} f(\eta, \nu, \omega, c),
\end{equation}
where the function $f$ is the dual-VQE cost function and is given by
\begin{multline}
f(\eta, \nu, \omega, c) \coloneqq \eta - c\big(\operatorname{Tr}[H^{2}] + \eta^{2}2^{n} + \nu^{2}\operatorname{Tr}[\omega^{2}]\\
- 2\eta\operatorname{Tr}[H] - 2\nu\operatorname{Tr}[H\omega]+2\eta\nu\big).
\end{multline}

\begin{figure*}
    \center
    \includegraphics[width=\textwidth]{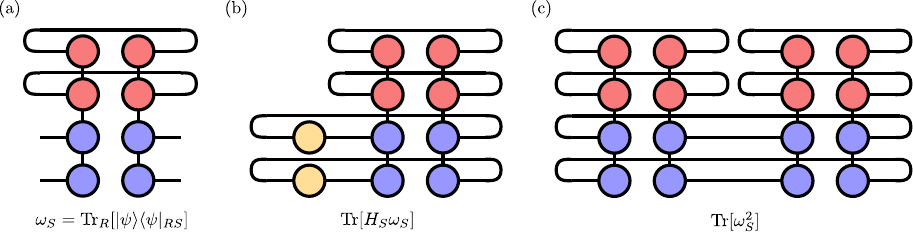}
    \caption{(a) Realizing mixed states using MPSs. The red tensors are the reference qubits, and the blue tensors are the system qubits. (b) Tensor network for estimating $\operatorname{Tr}[H\omega]$. The tensors in yellow represent the Pauli strings in the decomposition of~$H$. (c) Tensor network for estimating $\operatorname{Tr}[\omega^2]$.}
    \label{fig:mps-schematic-combined}
\end{figure*}

As discussed in Appendix~\ref{app:matrix-product-states}, the mixed state $\omega$ is realized as the partial trace of an MPS on a larger system. The MPS in this case represents the purification of the mixed state of interest. A schematic of this construction is given in Figure~\ref{fig:mps-schematic-combined}a). 

We begin with a randomly initialized MPS with $n_{S} + n_{R}$ nodes in the canonical form, where $n_{R}$ is the number of qubits in the reference system $R$, and $n_{S}$ is the number of qubits in the system $S$ of interest. Again, we pick $n_{R}$ to be at least as large as $n_{S}$ to ensure a full-rank mixed state on the system qubits. The two terms that depend on $\omega$ are $\operatorname{Tr}[H\omega]$ and $\operatorname{Tr}[\omega^2]$. These terms are estimated using the tensor networks in Figure~\ref{fig:mps-schematic-combined}b) and Figure~\ref{fig:mps-schematic-combined}c), respectively.

In our work, we use the {\tt quimb} package to realize matrix product states~\cite{Gray2018}. The package allows us to randomly initialize MPSs, contract specific indices, and compute gradients using automatic differentiation.

We begin all pretraining with a randomly initialized MPS where each entry of all the tensors are drawn from the normal distribution with zero mean and unit variance, and we use gradient ascent to train the parameters of the MPS to maximize the cost function in~\eqref{eq:dvqe-mps-costfn}.




\subsubsection{Convert final matrix product state to parametrized quantum circuit}

After pretraining, we convert the resulting MPS into a parametrized quantum circuit composed of one- and two-qubit gates. If the MPS has a maximum bond dimension of two ($\chi = 2$), it is exactly decomposable into a single layer of two-qubit unitaries~\cite{Ran2020a}. However, for an arbitrary MPS with $\chi > 2$, an exact decomposition into two-qubit gates is not generally feasible. This is because each MPS bond with dimension $\chi > 2$ corresponds to a unitary acting on $\lceil\log_2(\chi)\rceil + 1$ qubits, and it is not always possible to decompose any multi-qubit unitary exactly into a sequence of two-qubit gates.

To translate the final MPS into a parametrized quantum circuit, we make use of the algorithms put forth in~\cite{Rudolph_2024}. The building blocks of the different algorithms are called the Analytical Decomposition (AD) algorithm and Optimizing Decomposition (OD). We note that in this work and in prior works, the AD algorithm is also called the disentangling algorithm. 

For the sake of clarity and completeness, we provide a detailed description of both the AD and OD algorithms in Appendix~\ref{app:alg_AD_OD}.

\begin{remark}
    In~\cite{Rudolph_2024}, the OD algorithm consists of multiple iterations of training, where a unitary $U$ is replaced with a better unitary $U'$. To find this unitary $U'$, the algorithm uses the singular value decomposition (SVD) to find an optimal unitary $U_{\operatorname{new}}$, but it uses a small learning rate to prevent large updates. The update rule to replace unitary $U$ is as follows:
    \begin{equation} \label{eq:update_rule_former}
        U' = U (U^\dagger U_{\operatorname{new}})^\beta,
    \end{equation}
    where $\beta\in (0,1)$ is the learning rate, appropriately chosen beforehand.  
    In this work, we introduce a different update rule, given by
    \begin{equation}
    \label{eq:unitary-update-learning}
        U' = (1-\beta)\,U + \beta\,U_{\operatorname{new}}.
    \end{equation}
We note here that, as defined above, $U'$ is not unitary. To eliminate this issue, we perform an SVD of $U'$, obtaining $U' = W D V^{\dag}$ and we then set $U' \leftarrow W V^{\dag}$. In our numerical investigations, we observed that the update rule in \eqref{eq:unitary-update-learning} leads to faster convergence and better overall fidelity compared to using~\eqref{eq:update_rule_former}.
\end{remark}

In~\cite{Rudolph_2024}, the authors combined both the OD and AD algorithms in six different ways to translate the final trained MPS into a quantum circuit. From their work, we use three of the algorithms that they refer to as $D_{\operatorname{all}}$, $O_{\operatorname{all}}$, and $D_{\operatorname{all}}O_{\operatorname{all}}$. 
\begin{itemize}
    \item $D_{\operatorname{all}}$: Consists of the AD algorithm only. Multiple layers are created using the AD algorithm with no optimization.
    \item $O_{\operatorname{all}}$: Consists of the OD algorithm only. Multiple layers are randomly initalized and then optimized using the OD algorithm.
    \item $D_{\operatorname{all}}O_{\operatorname{all}}$: Begins with the AD algorithm as an initial circuit and then uses the OD algorithm to further improve fidelity.
\end{itemize}
We run all three variants above and finally pick the translated circuit with the highest fidelity. 

While the translated circuit is now composed of only two-qubit unitaries, it still needs to be cast into a form realizable on a quantum device. To that end, we utilize the KAK decomposition to decompose each two-qubit unitary into a product of parametrized rotation gates~\cite{KAK05}. Each two-qubit unitary contributes a total of 15 parameters.

\subsubsection{Train parametrized quantum circuit until convergence}
\label{subsub:quantum_training}

Lastly, we use the translated MPS to warm-start the quantum optimization. Before we begin the quantum training, we append additional layers of the parametrized ansatz that are trivially initialized (identity operations). Then, all the parameters, including those from the pretrained portion of the circuit, are trained. 

\subsubsection{Discussion of scalability}

\label{subsub:discussion-mps}

In Section~\ref{sec:Results}, we present simulated results for a small example problem of interest and find that the MPS pretraining procedure outlined above speeds up the quantum training significantly. Here, we appeal to prior numerical results~\cite{RMM+23} to argue that this performance advantage is expected to persist even as we increase the number of qubits beyond the reach of classical brute-force simulations.

The authors of Ref.~\cite{RMM+23} provided numerical evidence for the alleviation of barren plateaus by probing the gradient of the Kullback--Leibler (KL) divergence, for systems of up to twenty qubits. For the numerical examples considered, they found that random initializations led to the gradient variance decaying exponentially with system size (a key signature of the barren-plateau problem), whereas with MPS-pretrained states, the gradient variance did not decay significantly with either system size or the number of gate layers (see Figure~3 in~\cite{RMM+23} and the related discussion in the main text of~\cite{RMM+23}).

Furthermore, they extended these findings up to 100 qubits by estimating the gradient magnitude with an MPS-based quantum circuit simulator (for practical feasibility) again with respect to the KL divergence loss. They again found that random initialization led to the gradient vanishing with system size, whereas the gradient magnitude in the pretrained cases exhibited more stable behavior (see Figure~4 in~\cite{RMM+23} and the related discussion in the main text of~\cite{RMM+23}).

Overall, these prior findings suggest that the classical computational resources required to bypass the barren-plateau problem do not grow exponentially with system size. Nevertheless, we emphasize that strict scalability is not a prerequisite for achieving a practical advantage. Even if the classical resources needed to avoid barren plateaus become prohibitive at very large system sizes, this approach can still be useful. In this case, the pretraining algorithm may provide a good approximate initialization, which can then be refined on quantum hardware to yield improved performance.

\section{Results}

\label{sec:Results}

We now present the various results of simulations that test the performance of dual-VQE\ on the transverse-field, one-dimensional Ising model, described by the following Hamiltonian of $n$ qubits:
\begin{equation}\label{eq:ising_model}
    H_{\operatorname{TFI}}\coloneqq \sum_{i=1}^{n-1} J_i \sigma_{Z}^{i}\otimes\sigma_{Z}^{i+1} + \sum_{i=1}^{n} g_i\sigma_{X}^{i},
\end{equation}
where tensor products with identities are left implicit, $J_i$~is an interaction strength, and $g_i$ is related to the strength of the external field. In particular, we showcase our algorithm on the following example problem instance:
\begin{equation}
 H = \sum_{i=1}^{n-1} \sigma_{Z}^i \otimes \sigma_{Z}^{i+1} + \sum_{i=1}^n \sigma_{X}^i,
\end{equation}
for $n=2$ and $n=3$ qubits. We simulated the dual-VQE problem using three different ans\"atze -- the purification ansatz, the convex combination ansatz, and finally, MPS pretraining combined with a purification ansatz. We now present the details of the individual experiments and corresponding plots.

\begin{figure*}
\renewcommand{\arraystretch}{1.8}
\centering
\resizebox{.9\linewidth}{!}{

\begin{tabular}{P{0.13\linewidth}|c|c}
& \textbf{Purification ansatz} & \textbf{Convex-combination ansatz}
\vspace{-0.4cm}\\
\hline
\centering \textbf{Objective function}  & \vqedpsandwich & \vqedccasandwich \\
\hline
\centering\textbf{Error}  & \vqedperror & \vqedccaerror \\
\hline
\centering\textbf{Penalty value} & \vqedppenalty & \vqedccapenalty\\
\end{tabular}
}
\caption{Convergence of dual-VQE and comparison to standard VQE for a two-qubit example problem instance. The solid line shows the median value of the estimate, the shaded region represents the interquartile range, while the ground truth is marked by the dashed line. We were able to achieve an error of $10^{-2}$ after $20,000$ iterations of training.}
\label{fig:vqed-plots}
\end{figure*}

\subsection{Purification and convex combination without pretraining}

The results of experiments using the purification and convex combination ans\"atze are displayed in Figure~\ref{fig:vqed-plots}. The plots shown are for $n=2$ qubits. In these plots, solid lines depict results of both the standard VQE algorithm (with fixed learning rate of $0.005$ and gradient estimated by normalized Simultaneous Perturbation Stochastic Approximation (SPSA)) and our dual-VQE approach. The solid lines show median values across ten independent runs, and the shading is the interquartile range. The dashed line is the true value of the ground-state energy estimated via the SDP. 

Prior to training, we initialized the parameters $\eta$, $\nu$, and~$\theta$. We set the parameters $\eta$ and $\nu$ in~\eqref{eq:penalty-method-dual-VQE} to be equal to zero and one, respectively. We selected the components of the initial parameter vector $\theta$ for the state $\omega(\theta)$ uniformly at random from $[0,2\pi]$. The optimal state $\omega$ is almost always a full-rank mixed state. Thus, we used either a three-layer purification ansatz or a convex-combination ansatz with a two-layer quantum circuit Born machine and a two-layer unitary. The particular structure of the purification ansatz and the quantum circuit Born machine we use can be found in a Figure 2 of our companion paper 
\cite{Chen2023qslack}.

In addition to setting the initializations of the parameters of our optimization, we assigned specific values to the hyperparameters. We chose the penalty parameter $c$ to have a constant value of $10$ throughout training, and we implemented a decreasing learning rate scheme. We initialized the learning rate to a maximum value of $0.1$, and every $100$ iterations, we checked whether the objective function value decreased three times in a row over the past $300$ iterations. In this case, we reduced the learning rate by a factor of two. We repeated this process until a minimum value of $0.01$ was met.

Training crucially depends on estimating the gradient, which allows us to pick the next set of parameters. We used the SPSA algorithm~\cite{Spall1992}, which provides an unbiased estimate of the gradient at each step. We chose this method as the per-iteration cost of this algorithm is constant in the number of parameters, and we found that it works well in practice. If the norm of the gradient had a value greater than one, we normalized the gradient before performing the update step of gradient descent. We found that this prevents immediate divergence due to a potentially high initial penalty value. It also allowed us to use higher values of the learning rate than we would have been able to otherwise. Training concluded when an arbitrarily set maximum number of iterations was met.

The variation displayed in the graphs comes from three sources of randomness: the randomized initializations of all parametrized quantum circuits, randomness in the SPSA algorithm for gradient estimation, and shot noise over $10^{12}$ shots for every measurement of a quantum circuit. In our simulations, we were able to achieve an error of order $10^{-2}$ after $20,000$ iterations of training. These results give evidence that using VQE and dual-VQE allows us to bound the true value from both above and below, respectively, as desired.

\subsection{MPS pretraining with purification ansatz}

\begin{figure*}
    \centering
    \begin{subfigure}[t]{0.49\textwidth}
        \centering
        \includegraphics[width=\textwidth]{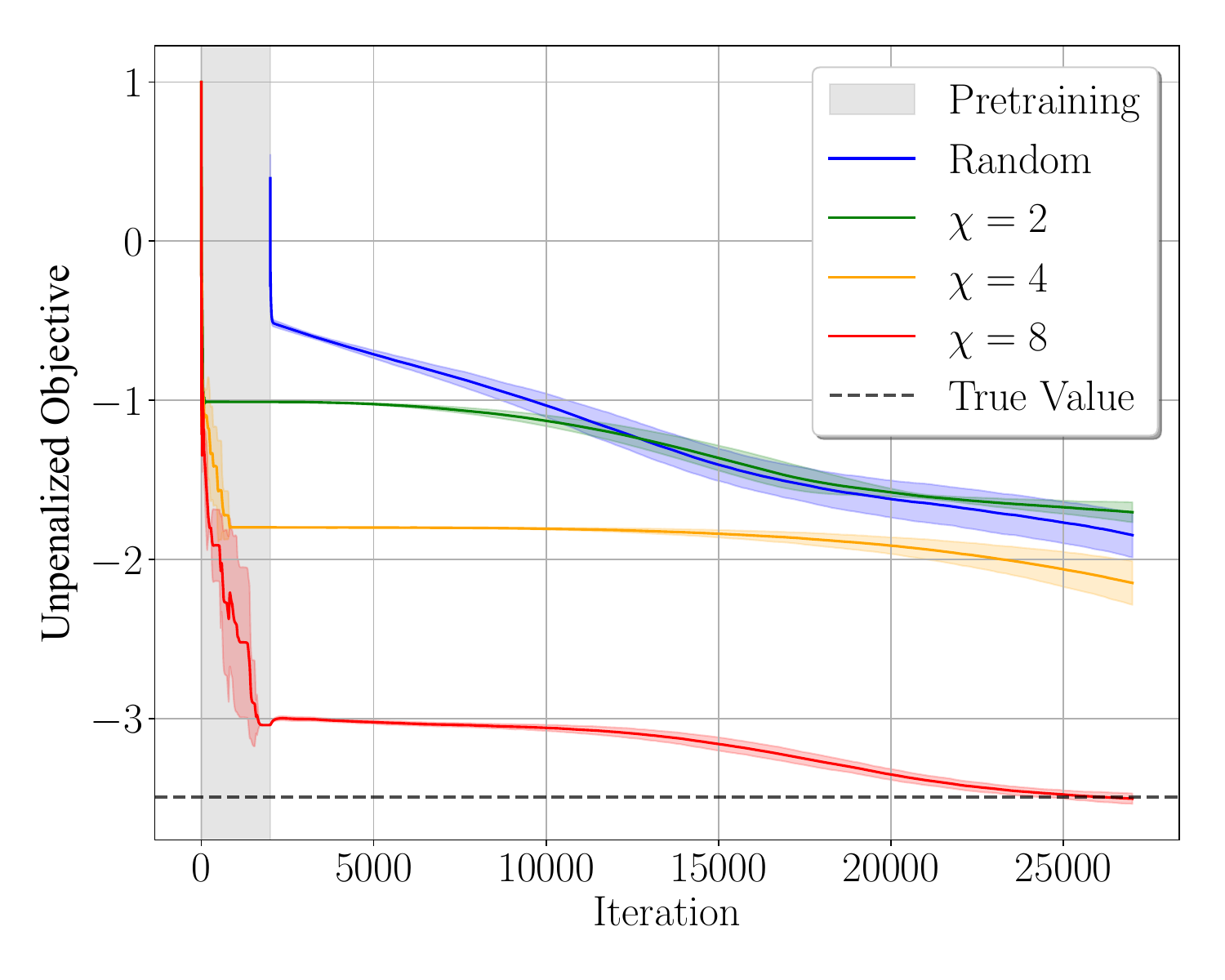}
    \end{subfigure}
    \hfill
    \begin{subfigure}[t]{0.49\textwidth}
        \centering
        \includegraphics[width=\textwidth]{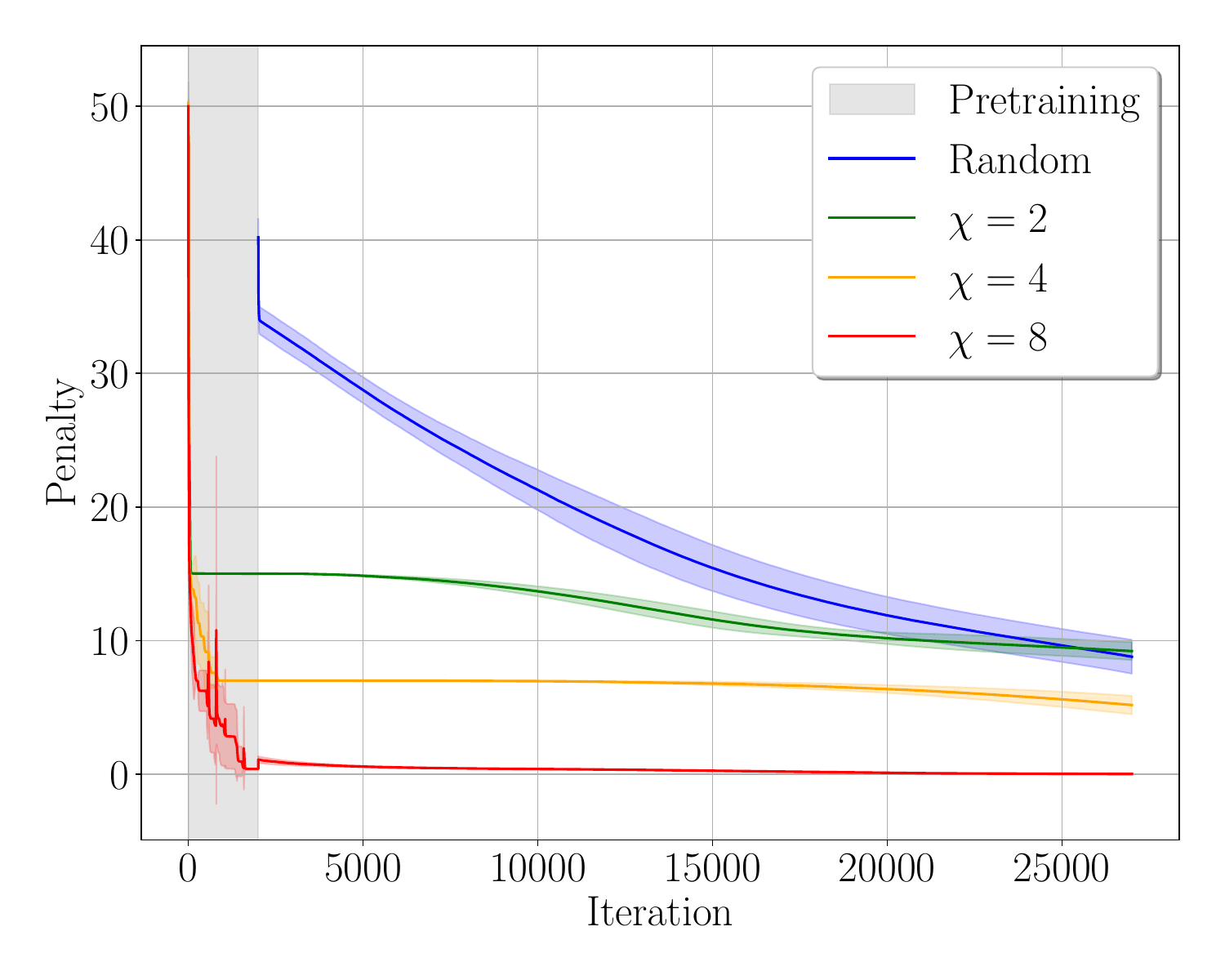}
    \end{subfigure}
    \caption{Unpenalized objective function value and penalty value for the Ising model problem using dual-VQE and MPS pretraining. After 2,000 iterations of pretraining followed by 25,000 iterations of quantum training, the true solution is reached when $\chi_{\max} = 8$, with a relative error of 0.5\%.} 
    \label{fig:pretraining-plots}
\end{figure*}

\begin{figure}
    \centering   
    \includegraphics[width=0.45\textwidth]{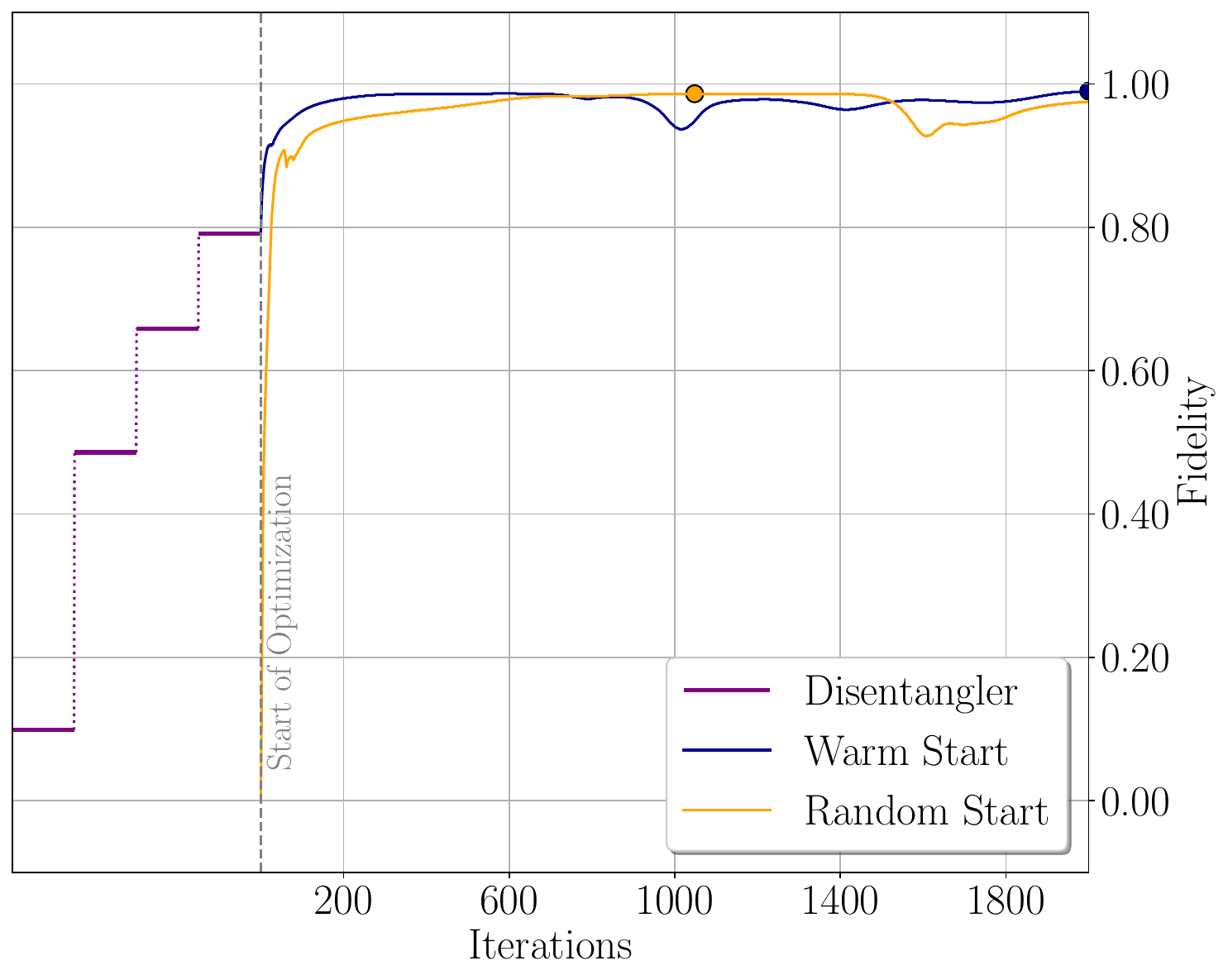}
    \caption{Fidelity of translation between the final MPS state after pretraining and the translated quantum circuit.}
    \label{fig:translation-fidelity}
\end{figure}

The results of the different experiments where we used MPS pretraining are given in Figure~\ref{fig:pretraining-plots}. The plots shown are for $n=3$ qubits. To demonstrate the effectiveness of pretraining, we used three different max bond dimension values $\chi_{\operatorname{max}} = \{2, 4, 8\}$ and compared it with the no-pretraining case. In each of the above cases, we conducted five runs and report the mean and standard deviation. 

For the three pretraining runs, with $\chi_{\operatorname{max}} = \{2, 4, 8\}$, we used a randomly initialized MPS on six qubits to realize a full-rank mixed state state on three qubits. We set the initial penalty parameter to be $c=30$ and kept it constant for the pretraining phase. We used gradient descent to train for $2000$ iterations, labeled by \textbf{Pretraining} in Figure~\ref{fig:pretraining-plots}.

Following the pretraining, we translated the MPS into a quantum circuit. We used the AD decomposition to decompose the MPS into three layers of two-qubit unitaries, labeled as \textbf{Disentangler} in Figure~\ref{fig:translation-fidelity}.
We plot the fidelity of translation, using the AD and OD algorithms (defined in Appendix~\ref{app:alg_AD_OD}), in Figure~\ref{fig:translation-fidelity}. 

The circuit we obtained then serves as the starting point for one of the OD algorithm runs. For this run, labeled as \textbf{Warm Start}, we used 2000 iterations and a constant learning rate of $\beta = 0.2$ (see~\eqref{eq:unitary-update-learning}). We also ran the OD algorithm on a randomly initialized three layer circuit and trained for 2000 iterations, labeled as \textbf{Random Start}. Finally, we picked the best translated circuit. For $\chi_{\max} = 8$, the translation fidelity was around~$99.5\%$.

Lastly, we added a single layer of parametrized gates to the translated circuits, as described in Section~\ref{subsub:quantum_training}, and all the parameters are trained using a combination of a finite difference method and the SPSA algorithm. For the parameters $\eta$ and $\nu$, we used a finite difference method with a perturbation of $10^{-6}$. For the circuit parameters, we used the SPSA algorithm since it provides an unbiased estimate of the gradient with a runtime complexity that is independent of the number of parameters. For $\eta$ and $\nu$, we initialized the learning rate, denoted by $\beta_{\eta, \nu}$, to be $10^{-3}$ and for the circuit parameters, we initialized the learning rate, denoted by $\beta_\omega$, to be $3 \times 10^{-3}$. Every 200 iterations, we modified the learning rates and the penalty parameter $c$. For the learning rates, we calculated the slope of the last 200 iterations. If the slope was negative, we set 
\begin{align}
    \beta_{\omega} &= \max(0.9\beta_{\omega}, \beta_{\min}), \\
    \beta_{\eta, \nu} &= \max(0.9\beta_{\eta, \nu}, 3\beta_{\min}).
\end{align}
On the other hand, if the slope was positive, we set 
\begin{align}
    \beta_{\omega} &= \min(1.05\beta_{\omega}, \beta_{\max}), \\
    \beta_{\eta, \nu} &= \min(1.02\beta_{\eta, \nu}, 3\beta_{max}).
\end{align}
In the equations above, $\beta_{\max} = 10^{-3}$ and  $\beta_{\min} = 10^{-4}$.

Similarly, we updated the penalty parameter by calculating the slope of the penalty of the last 200 iterations. If the slope was greater than $5 \times 10^{-4}$, we updated $c$  to be $0.9c$,
and if the slope was between $0.0$ and $5 \times 10^{-4}$, we set $c = \min(1.04c, 40)$. This latter case occurs when the penalty is stagnated, and usually occurs when the minimization has come close to a minima for the current $c$ value. So, we increased the $c$ value and enforced the constraint more strongly. These penalty updates deviate from the presentation in Section~\ref{sec:alg-cons}, but we have found that they work well in practice for the problems we considered.

We ran the quantum training for 25,000 iterations. The plots show that pretraining even for a small number of iterations speeds up the quantum training significantly. Furthermore, we see that increasing the bond dimension further increases the advantage. For $\chi_{\max} = 8$, we see that the quantum training reaches the true solution with a relative error of $0.5\%$. 

\section{Discussion and conclusion}
\label{sec:discussion}

In conclusion, we proposed dual-VQE as a variational quantum algorithm for lower bounding the ground-state energy of a Hamiltonian. This lower bound can be compared with
the upper bound obtained from the traditional VQE\ method, and the two bounds
serve as quality checks on each other that ideally determine an interval
containing the optimal ground-state energy. As discussed above, the quantum
computational resources required for evaluating the objective function of dual-VQE\ are no greater than those needed for VQE's objective function when using the convex combination ansatz in conjunction with a classical generative model.

While Theorem~\ref{thm:dVQE-lower-bound} guarantees that dual-VQE provides a rigorous lower bound on the ground-state energy in the limit $c \rightarrow \infty$, for finite $c$, the value provided by dual-VQE is technically an estimate rather than a strict bound. However, our numerical simulations give evidence, for the examples considered, that VQE and dual-VQE do indeed bound the optimal ground-state energy from above and below, respectively.  Nonetheless, it would be valuable to investigate whether it is possible to obtain bounds on the degree to which deviations are possible for finite $c$. This would allow us to certify the reliability of dual-VQE and serve as a guide to an optimum choice for $c$.



For future work, it would certainly be of interest to scale up dual-VQE to much larger systems beyond the reach of brute-force classical simulation and to compare the results of VQE and dual-VQE with each other. We also think it would be worthwhile to pursue other methods, besides the penalty method, for performing the optimization in dual-VQE. A promising approach would be to combine a variational approach with an interior-point method, but for this approach, it seems we would need an efficient method, suitable for near-term quantum computers, for estimating the logarithm of the determinant of a matrix on a quantum computer. The quantum algorithms put forward in~\cite{Zhao2019,luongo2024,giovannetti2025} could be helpful for this purpose. 

Finally, matrix product pretrainers are just one class of pretrainers that can help find advantageous initial points. Exploring other pretraining methods, including classical neural networks, may lead to more efficient optimization schemes. 

\textit{Data availability statement}---All source codes used to run the simulations and generate the figures are available on Zenodo \cite{WCZ+23}.

\begin{acknowledgments}
We thank Paul Alsing, Ziv Goldfeld, Daniel Koch, Saahil Patel, and Manuel
S.~Rudolph for helpful discussions. JC and HW acknowledge support from the
Engineering Learning Initiative in Cornell University's College of
Engineering. ZH acknowledges support from the Sandoz Family Foundation Monique
de Meuron program for Academic Promotion. IL, TN, DP, SR, KW, and MMW acknowledge
support from the School of Electrical and Computer Engineering at Cornell
University. TN, DP, SR, and MMW acknowledge support from the National Science
Foundation under Grant No.~2315398. DP, SR, and MMW acknowledge support from
AFRL under agreement no.~FA8750-23-2-0031.

This material is based on research
sponsored by Air Force Research Laboratory under agreement number
FA8750-23-2-0031. The U.S.~Government is authorized to reproduced and
distribute reprints for Governmental purposes notwithstanding any copyright
notation thereon. The views and conclusions contained herein are those of the
authors and should not be interpreted as necessarily representing the official
policies or endorsements, either expressed or implied, of Air Force Research
Laboratory or the U.S.~Government. 

This research was conducted with support from the Cornell University Center for Advanced Computing, which receives funding from Cornell University, the National Science Foundation, and members of its Partner Program.
\end{acknowledgments}










\bibliography{Ref-dual-VQE}

\appendix

\section{Matrix product states}

\label{app:matrix-product-states}

An $n$-qubit pure state is a vector that belongs to a complex Hilbert space of dimension $2^n$, thus requiring $2^n$ complex numbers to fully specify it. Furthermore, unitary operations acting on this pure state are, in general, of dimension $2^n \times 2^n$ and thus require both an exponential amount of memory and computation to implement. In tensor-network notation, such a vector is generally represented by a sphere with a single leg to specify the index~\cite{Berezutskii2025TensorNetworks}.

\begin{figure}
    \center
    \includegraphics[width=0.9\columnwidth]{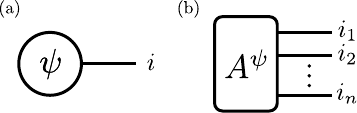}
    \caption{(a) Representation of a pure state in tensor-network notation. (b) Alternate representation of a pure state in  tensor-network notation.}
    \label{fig:pure-state-reps}
\end{figure}

For example, an $n$-qubit pure state $\vert \psi \rangle$ can be expanded as 
\begin{equation}
    \vert \psi \rangle = \sum_{i=0}^{2^n-1} \psi_i \vert i \rangle,
\end{equation}
where $\psi_i \in \mathbb{C}$ for all $i\in \{0,\ldots, 2^n-1\}$, subject to $\sum_{i=0}^{2^n-1} \left | \psi_i \right|^2=1$,
and the corresponding figure is given in Figure~\ref{fig:pure-state-reps}a). Here, $\psi$ represents the rank-$1$ tensor that contains the components of $\vert \psi \rangle$. 

An alternate way of representing the same state involves using an $n$-legged tensor with each leg representing a single-qubit system. More concretely, the $n$-qubit pure state $\vert \psi \rangle$ can be expanded as 
\begin{equation}
    \vert \psi \rangle = \sum_{i_1, i_2, \ldots, i_n = 0}^{1} (A^\psi)_{i_1 i_2 \cdots i_n} \vert i_1 i_2 \cdots i_n \rangle,
\end{equation}
and the corresponding figure is given in Figure~\ref{fig:pure-state-reps}b). Here, $A^\psi$ represents the rank-$n$ tensor that contains the components of $\vert \psi \rangle$, using the following mapping:
\begin{equation}
    (A^\psi)_{i_1 i_2 \cdots i_n} = \psi_i,
\end{equation}
where $i_1 i_2 \cdots i_n$ is the binary representation of $i$. Both representations above are equivalent to each other, and they represent no approximations or simplifications. 

To manage the computational and spatial complexity of these large tensors, they can be factorized into a network of lower-order tensors. This factorization enables efficient storage and computation, depending on the resulting axis dimensions. One such factorization of this large tensor is in the form of a matrix product state. 

\begin{figure}[ht]
    \center
    \vspace{0.5em}
    \includegraphics[width=0.3\columnwidth]{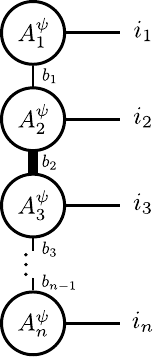}
    \caption{MPS representation of the state $\vert \psi \rangle$.}
    \label{fig:mps-schematic}
\end{figure}

Matrix product states have a linear architecture, with each tensor being connected to their nearest neighbors. For example, the tensor $A^\psi$ can be factorized into a linear chain of smaller tensors, as seen in Figure~\ref{fig:mps-schematic}.

\begin{figure*}
    \center
    \includegraphics[width=0.75\textwidth]{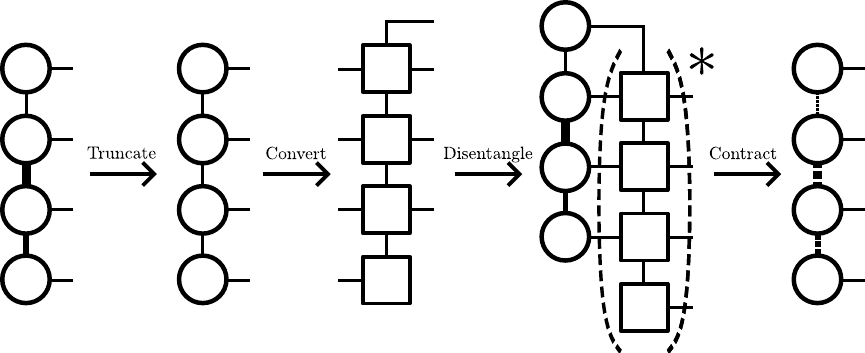}
    \caption{Analytical Decomposition algorithm. Reproduced from~\cite{Rudolph_2024}. The steps are described in Algorithm \ref{alg:analytic-decomposition}.}
    \label{fig:mps-AD-algorithm}
\end{figure*}

Mathematically, this is expanded as 
\begin{align}
    \vert \psi \rangle &= (A^\psi)_{i_1 i_2 \cdots i_n} \vert i_1 i_2 \cdots i_n \rangle \notag \\
    &= (A^\psi_1)^{b_1}_{i_1} (A^\psi_2)^{b_2}_{b_1 i_2} \cdots (A^\psi_n)^{b_{n-1}}_{i_n} \vert i_1 i_2 \cdots i_n \rangle,
\end{align}
where we use the Einstein repeated index summation convention. The indices labeled by $b_k$ are called bonds, and their dimension is called the bond dimension of the MPS. Note that, in general, the bonds can be of different dimension, in which case, the largest bond is quoted as the bond dimension of the MPS and is denoted by $\chi$. In addition to the bond dimension, the size of the output indices is called the physical dimension. In this work, since we are working with qubits, we fix the physical dimension to be $2$. Thus, the number of parameters of the entire MPS is given by $2n\chi^2$, as opposed to $2^n$ for the full state vector.

Matrix product states allow us to efficiently represent general pure states with the bond dimension controlling the level of compression. To fully express the entire set of pure states, the bond dimension needs to grow exponentially in $n$. Thus, the chosen $\chi$ can be increased to explore larger sectors of the entire Hilbert space~\cite{Oru14}. The physical interpretation of the bond dimension is that it is a measure of the entanglement present in the state. Thus, MPSs with small bond dimension $\chi$ represent states with low entanglement. When used in optimization tasks, tuning the bond dimension provides for a method to smoothly increase the entanglement present in an ansatz, exploring more entangled states in the Hilbert space.

The advantages of using tensor networks do not end here. Matrix product operators generalize unitary operations and can be implemented using tensor contractions in a runtime polynomial in the bond dimension. For states with entanglement greater than the max bond dimension can exactly reproduce, the singular value decomposition may be used to find a low bond dimension MPS approximation~\cite{SCHOLLWOCK201196}.

Just as mixed states in a Hilbert space can be represented by a pure state in a larger Hilbert space, so too can larger rank MPSs be used to represent mixed states on a smaller rank. For example, consider a mixed state $\omega_S$ that has a purification $\vert \psi\rangle_{RS}$. If $\vert \psi \rangle_{RS}$ has an MPS representation, then $\omega_S$ can be realized as a matrix product operator (MPO) using this MPS, as depicted in Figure~\ref{fig:mps-schematic-combined}a).

\begin{figure*}
    \center
    \includegraphics[width=0.85\textwidth]{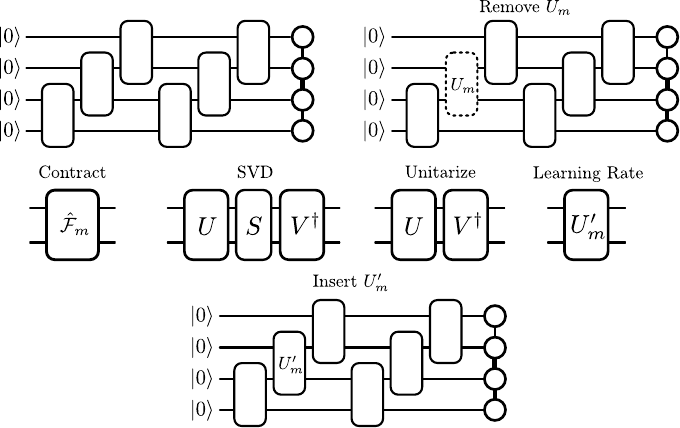}
    \caption{Optimizing Decomposition algorithm. Reproduced from~\cite{Rudolph_2024}. The steps are described in Algorithm \ref{alg:optimizing-decomposition}.}
    \label{fig:mps-OD-algorithm}
\end{figure*}

\section{Algorithms from RCMAP23}

\label{app:alg_AD_OD}

In this appendix, we review some algorithms studied in~\cite{Rudolph_2024}. The algorithms are a combination of the Analytic Decomposition (AD) and the Optimizing Decomposition (OD) algorithms. The AD algorithm uses MPS truncation to find the optimal layer of two-qubit unitaries to disentangle the current MPS, i.e., to reduce its bond dimension. The process is repeated until a fixed iteration count is met or the fidelity reaches a threshold. On the other hand, the OD algorithm begins with a random circuit of two-qubit unitaries and systematically replaces each one with a more optimal unitary. For the sake of clarity and completeness, we provide a detailed description of both the AD and OD algorithms below.

\subsubsection{Analytic Decomposition (AD)}

The Analytic Decomposition (AD) algorithm was proposed as a method for decomposing an MPS of arbitrary bond dimension $\chi$ into multiple linear layers of two-qubit unitaries~\cite{Ran2020a}. The AD algorithm is iterative, introducing a new layer of two-qubit unitaries in each iteration until a stopping criterion is met. The stopping criterion is usually a combination of an upper bound on the number of layers, or the fidelity between the MPS and the translated quantum circuit. These linear layers have a staircase arrangement. Let us denote such a layer of two-qubit unitaries as $L[U]$.

The algorithm is aimed at maximizing the fidelity defined by 
\begin{align}
    f(\{L[U]^{(k)}\}_{k=1}^K) &= \vert \langle 0^{\otimes N} \vert \prod\limits_{k=1}^K {L[U]^{(k)}}^\dagger \vert \psi_{\chi_{\max}} \rangle \vert^2 \notag \\
    &= \vert \langle 0^{\otimes N} \vert {L[U]^{(K)}}^\dagger \cdots {L[U]^{(1)}}^\dagger \vert \psi_{\chi_{\max}} \rangle \vert^2 \notag \\
    &= \vert \langle \psi^{(K)}_{\text{QC}} \vert \psi_{\chi_{\max}} \rangle \vert^2.
\end{align}
At each iteration, one layer of two-qubit unitaries is created that disentangles the truncated MPS to the all-zeros state. Then, using this layer of unitaries, the current MPS is disentangled, and then contracted, to get the MPS with lower bond dimension at the next iteration. The steps of the algorithm can be found in Algorithm~\ref{alg:analytic-decomposition} and a schematic can be found in Figure~\ref{fig:mps-AD-algorithm}. 

\begin{algorithm}[H]
\caption{Analytic Decomposition (AD)}
\label{alg:analytic-decomposition}
\begin{algorithmic}[1]
\vspace{0.2em}
\item[] \hspace*{-\algorithmicindent}
\begin{minipage}{\dimexpr\linewidth-\algorithmicindent\relax}
\textbf{Input:}\vspace{-0.5em}
\begin{itemize}\setlength\itemsep{-0.2em}
    \item MPS $\psi_{\chi_{\max}}$
    \item Maximum layers $K$
    \item Target fidelity $\hat{f}$
\end{itemize}
\end{minipage}
\item[] \hspace*{-\algorithmicindent} \textbf{Output:} Quantum Circuit layers $\Pi_{k=K}^1 L[U]^{(k)}$
\vspace{0.5em}
\STATE Set $k = 1$.
\STATE Set $|\psi^{0}\rangle = |\psi_{\chi_{\max}}\rangle$.

\WHILE{$k \leq K$ and $\vert \langle 0^{\otimes N} \vert \psi^{k} \rangle \vert^2 < \hat{f}$}
    \STATE Truncate $\psi^k$ to $\psi^k_{\chi=2}$ via SVD.
    \STATE Convert $\psi^k_{\chi=2}$ to $L[U]^{(k+1)}$.
    \STATE Append $L[U]^{(k+1)}$ to get $\vert \psi^{(k+1)} \rangle = {L[U]^{(k+1)}}^\dagger \vert \psi^{(k)} \rangle$.
    \STATE Increment $k = k+1$.
\ENDWHILE
\end{algorithmic}
\end{algorithm}

\subsubsection{Optimizing Decomposition (OD) }

The Optimizing Decomposition (OD) algorithm begins with a random layered circuit made of two-qubit unitaries and iteratively replaces each two-qubit unitary with another that increases the fidelity of the prepared state with the target MPS. The steps of the algorithm can be found in Algorithm~\ref{alg:optimizing-decomposition} and a schematic can be found in Figure~\ref{fig:mps-OD-algorithm}. 

In each iteration, the environment tensor $\hat{\mathcal{F}}_m$ is calculated by first removing the unitary $U_m$ and contracting all other indices. This results in a rank-$4$ tensor, which is not necessarily unitary. It turns out that $\hat{\mathcal{F}}_m$ is the locally optimal operator to insert in place of $U_m$ to increase the fidelity. However, since it is not generally unitary, we instead find the closest unitary to $\hat{\mathcal{F}}_m$ by applying the SVD and discarding the singular values. We call this unitary $U_{\operatorname{new}, m}$. However, replacing $U_m$ with $U_{\operatorname{new}, m}$ is a strong update. Instead, we use a learning rate to smoothly interpolate between the two extremes. In this work, we use a linear learning rate 
\begin{equation}
    U'_m = (1-\beta)\,U_m + \beta\,U_{\operatorname{new}, m},
\end{equation}
where $\beta\in(0,1)$ is the learning rate appropriately chosen beforehand. We note here that, as defined above, $U'_m$ is not unitary. To mitigate this, we perform an SVD and discard the singular values again. \\


\begin{algorithm}[H]
\caption{Optimizing Decomposition (OD)}
\label{alg:optimizing-decomposition}
\begin{algorithmic}[1]
\vspace{0.2em}
\item[] \hspace*{-\algorithmicindent}
\begin{minipage}{\dimexpr\linewidth-\algorithmicindent\relax}
\textbf{Input:}\vspace{-0.5em}
\begin{itemize}\setlength\itemsep{-0.2em}
    \item Initial quantum circuit $\prod_{m=1}^{M} U_{m, 0}$
    \item Number of sweeps $T$
    \item Target fidelity $\hat{f}$
    \item Learning rate $r \in [0, 1]$
\end{itemize}
\end{minipage}
\item[] \hspace*{-\algorithmicindent} \textbf{Output:} Optimized quantum circuit $\prod_{m=1}^M U_m$
\vspace{0.5em}
\STATE Set $t = 1$.
\WHILE{$t \leq T$ and $\vert \langle 0^{\otimes N} \vert \prod_{m=M}^{1}U_m^\dagger \vert \psi_{\chi_{\max}} \rangle \vert^2 < \hat{f}$}
    \FOR{$m$ in $1 \ldots M$}
        \STATE Calculate $\hat{\mathcal{F}}_m$.
        \STATE Split using SVD $\hat{\mathcal{F}} = \mathcal{U}\mathcal{S}\mathcal{V}^\dagger$.
        \STATE Define new unitary $U_{\text{new,}m} = \mathcal{U}\mathcal{V}^\dagger$
        \STATE Apply learning rate to get new $U_m$.
    \ENDFOR
    \STATE Increment $t = t + 1$.
\ENDWHILE
\end{algorithmic}
\end{algorithm}

\end{document}